\documentclass{llncs}

\usepackage{makeidx}
\usepackage{graphicx}
\usepackage{subcaption}
\usepackage{enumerate}
\usepackage[noadjust]{cite}
\usepackage[hidelinks, bookmarks=false, draft]{hyperref}
\usepackage{multirow}
\usepackage{booktabs}
\usepackage{dblfloatfix}
\usepackage{marvosym}
\usepackage{enumitem}
\usepackage{hhline}
\usepackage[multiple]{footmisc}
\usepackage{multirow}
\usepackage{array}
\usepackage[table]{xcolor}
\newcolumntype{P}[1]{>{\centering\arraybackslash}p{#1}}

\usepackage[utf8]{inputenc}
\usepackage{amsmath}
\usepackage{url}
\usepackage{mdframed}
\usepackage{amssymb}
\usepackage{subcaption}
\usepackage{caption}
\usepackage{csquotes}

\mdfsetup{skipabove=3pt,skipbelow=3pt}
\newcommand{\rulesep}{\unskip\ \vrule\ }

\begin{document}


\title{Software Professionals' Attitudes towards Video as a Medium in Requirements Engineering}


\titlerunning{Video as a Medium in Requirements Engineering}

\author{Oliver Karras \Letter}

\authorrunning{Oliver Karras}

\institute{Software Engineering Group\\ Leibniz Universit\"at Hannover 
\\30167 Hannover, Germany\\
\email{Email: oliver.karras@inf.uni-hannover.de}}


\maketitle

\begin{abstract}
In requirements engineering (RE), knowledge is mainly communicated via written specifications. This practice is cumbersome due to its low communication richness and effectiveness. In contrast, videos can transfer knowledge more richly and effectively. However, video is still a neglected medium in RE.
We investigate if software professionals perceive video as a medium that can contribute to RE. We focus on their attitudes towards video as a medium in RE including its strengths, weaknesses, opportunities, and threats.
We conducted a survey to explore these attitudes with a questionnaire. 64 out of 106 software professionals completed the survey.
The respondents' overall attitude towards video is positive. 59 of them stated that video has the potential to improve RE. However, 34 respondents also mentioned threats of videos for RE. We identified the strengths, weaknesses, opportunities, and threats of videos for RE from the point of view of software professionals.
Video is a medium with a neglected potential. Software professionals do not fundamentally reject videos in RE. Despite the strengths and opportunities of video, the stated weaknesses and threats impede its application. Based on our findings, we conclude that software professionals need guidance on how to produce and use videos for visual communication to take full advantage of the currently neglected potential.

\end{abstract}

\section{Introduction}
\label{sec:introduction}
One of the most widely used documentation options to convey stakeholders' needs is a written specification as suggested by standards such as ISO/IEC/IEEE $29148$:$2011$ \cite{ISO29148.2011}. However, the supposedly simple handover of a written specification insufficiently supports the rich information and knowledge transfer which is necessary to develop an acceptable system \cite{Fricker.2010, Klunder.2016}. Abad et al. \cite{Abad.2016} found the need for improving requirements communication by exceeding pictorial representations in written specifications. The authors proposed to invest more efforts in addressing interactive visualizations such as storytelling, for example with videos \cite{Abad.2016}. In the last $35$ years, several researchers \cite{Feeney.1983, Creighton., Brill.2010, Karras.2016b, Karras.2017a} proposed approaches that use videos in RE to support requirements communication. Despite all this research, video is still not an established documentation option in terms of RE best practice \cite{Fricker.2015c}.
In our recently published position paper \cite{Karras.2018}, we discussed video production in RE. In accordance with the aforementioned researchers \cite{Abad.2016, Feeney.1983, Creighton., Brill.2010, Karras.2016b, Karras.2017a}, we also concluded that software professionals could enrich their communication and RE abilities if they knew what constitutes a good video for visual communication. As future work, we proposed to develop a quality model for videos to encourage and enable software professionals to produce effective videos on their own \cite{Karras.2018}.

However, our future work and probably the existing approaches \cite{Feeney.1983, Creighton., Brill.2010, Karras.2016b, Karras.2017a} are based on the assumption that software professionals perceive video as a medium that can contribute to RE. In this paper, we investigate this assumption by conducting an explorative survey focusing on the following research question:

\begin{mdframed}
	\begin{itemize}[leftmargin=-2.5mm]
		\item[] \textbf{Research question:}
		
		What are software professionals' attitudes towards video as a medium in RE?
	\end{itemize}
\end{mdframed}

Based on the attitudes, we expect to achieve insights that either substantiate or refute the assumption. By answering this research question, we can understand if software professionals fundamentally reject video as a medium in RE. Such a rejection would be reflected in a negative attitude including the mention of weaknesses and threats of videos. Otherwise, we assume a neutral or even positive attitude towards video including the mention of strengths and opportunities. Therefore, this information provides insights into the current perception of videos in RE by software professionals. We contribute the following insights.

Software professionals have generally a positive attitude towards video as a medium in RE. Although $59$ respondents state that video can improve RE, this medium has a neglected potential. The identified strengths and opportunities of videos such as \textit{richness}, \textit{simplicity}, improved \textit{communication}, and improved \textit{understanding} indicate the benefits of video as a powerful and simple documentation option for communication. However, the mentioned weaknesses and threats such as high \textit{effort}, technical \textit{constraints}, \textit{misuse}, and \textit{intimidation} impede the application of videos in RE. Furthermore, they indicate a lack of knowledge by software professionals on how to produce and use good videos.

\section{Documentation for Communication: A Challenge of RE}
\label{sec:related-work}
Different studies investigated RE practices in terms of documentation and communication \cite{Al-Rawas.1996, Fricker.2015c, Lethbridge.2003, Carter.2009, Abad.2016}. All of them indicate that a written specification is ($1$) the most common medium for requirements communication and ($2$) a crucial RE challenge due to its low communication richness and effectiveness.

In a field study, Al-Raws and Easterbrook \cite{Al-Rawas.1996} found that written specifications insufficiently support communication due to the inherent restrictions of available notations. They conclude that specifications need to be enriched in order to turn them into an effective means of communication.
Fricker et al. \cite{Fricker.2015c} also conducted a survey on RE practices. Their results show that all applied and established documentation notations consists only of pictorial or textual representations.
Lethbridge et al. \cite{Lethbridge.2003} performed a study on the use of documentation. Their findings indicate that software professionals often perceive documentation as too complex. The authors conclude the necessity to focus on power and simplicity of documentation to increase its relevance.
Carter and Karatsolis \cite{Carter.2009} reported lessons learned from developing a robust documentation. Based on their experiences, they suggest to include multimedia documentation, such as videos, in RE. The authors believe that adding such multimedia documentations to the RE palette of notations can produce a significant value.
Abad et al. \cite{Abad.2016} conducted a systematic literature review on visualization in RE. One of their key findings is the need for a better support of requirements communication that exceeds pictorial representations in written specifications.

All previously mentioned studies indicate a still existing need for improving documentation for communication in RE. Several researchers \cite{Feeney.1983, Creighton., Brill.2010, Karras.2016b, Karras.2017a} addressed this problem by focusing on the use of video as a documentation option in RE. They followed the line-of-thought of Lethbridge et al. \cite{Lethbridge.2003} as well as Carter and Karatsolis \cite{Carter.2009}. Despite its communication richness and effectiveness, video is still not an established documentation option in RE. Therefore, we conducted a survey on software professionals' \textit{attitudes towards video as a medium in RE}.

\section{Survey -- Video as a Medium in RE}
\label{sec:survey----video-as-a-medium-in-re}
We aligned the survey design by following the steps and guidelines for carrying out a questionnaire survey as proposed by Robson and McCartan \cite[p. 244 ff.]{Robson.2016}.

\textbf{Design:} We iteratively refined the questionnaire design consisting of $6$ closed (demographics and attitude) and $4$ open-ended (strengths, weaknesses, opportunities, and threats) questions. We performed the initial testing by using the checklist provided by Baum et al. \cite{Baum.2017} to review every single question. This was followed by $5$ rounds of pre-tests. In each pre-test, a software professional completed the survey and we discussed how the questionnaire could be improved.

\textbf{Data Collection:} In late $2017$, we conducted the survey
implemented in LimeSurvey. We relied on a number of communication channels to reach suitable participants, e.g. LinkedIn, ResearchGate, a mailing list of a German RE professionals group, and advertisement at the \textit{25th IEEE International Requirements Engineering Conference}. Our target population included practitioners and researchers since both groups have relevant attitudes towards video as a medium in RE. While practitioners report an industrial, project-oriented point of view, researchers state a scientific, project-oriented one.

\textbf{Analysis:} We analyzed the open-ended questions with manual coding \cite{Saldana.2015}. This is a qualitative data analysis consisting of two consecutive coding cycles of which each can be repeated iteratively. The first cycle includes the initial coding of the data. The second cycle focuses on classifying, abstracting, and conceptualizing categories from the coded data. In the first cycle, we applied \textit{in vivo coding} which assigns a word or phrase found in a response as a code to the respective data. In the second cycle, we performed \textit{pattern coding} which groups the coded data into categories. We iterated three times through each cycle.

\subsection{Survey Results}
\label{sec:results}
\textbf{Demography:} 
The respondents worked in $11$ countries: $40$ in Germany, $16$ in other European countries, $6$ in North America, and $2$ in Asia including the Middle East. Of $64$ respondents $34$ were from industry and $30$ were from academia. $8$ practitioners stated their job as \textit{requirements engineer}, $7$ as \textit{project manager}, $5$ as \textit{developer}, $2$ as \textit{software architect}, and $12$ as other business roles only mentioned once. The researchers stated mainly two research areas: $16$ times \textit{requirements engineering} and $10$ times \textit{software engineering}. $4$ respondents mentioned other research areas in computer science which were only mentioned once. On average, the practitioners had $9.2$ years of experience and the researchers $7.4$ years.

\textbf{Attitudes towards Video:}
Of the $64$ respondents, $38$ had a positive, $25$ a neutral and $1$ a negative attitude towards video as a medium in RE. $59$ respondents stated that videos have the potential to improve RE. $34$ respondents mentioned threats of video for RE. \tablename{ \ref{tbl:t1}} summarizes the previously described findings. All respondents stated at least one strength and one weakness of video.

\vspace{-0.9cm}
\begin{table}[]
	\renewcommand{\arraystretch}{1.2}
	\centering
	\caption{Video as a medium in RE: Attitudes, potential, and threats}
	\label{tbl:t1}
	\begin{tabular}{P{1.3cm}|P{1.3cm}|P{1.3cm}|P{1.3cm}||P{1.4cm}|P{1.4cm}||P{1.3cm}|P{1.3cm}|}		
		\cline{2-8}
		\multicolumn{1}{c|}{\multirow{2}{*}{}} & \multicolumn{3}{c||}{\textbf{Attitude towards Videos}} & \multicolumn{2}{c||}{\textbf{Potential for RE?}} & \multicolumn{2}{c|}{\textbf{Threats for RE?}} \\ \cline{2-8} 
		\multicolumn{1}{c|}{} & Positive & Neutral & Negative & Yes & No & Yes & No \\ \hline
		\multicolumn{1}{|c|}{Researcher} & 21 & 9 & 0 & 28 & 2 & 17 & 13 \\ \hline
		\multicolumn{1}{|c|}{Practitioner} & 17 & 16 & 1 & 31 & 3 & 17 & 17 \\ \hline \hline
		\multicolumn{1}{|c|}{\cellcolor[HTML]{C0C0C0}\textbf{Total}} & \cellcolor[HTML]{C0C0C0}38 & \cellcolor[HTML]{C0C0C0}25 & \cellcolor[HTML]{C0C0C0}1 & \cellcolor[HTML]{C0C0C0}59 & \cellcolor[HTML]{C0C0C0}5 & \cellcolor[HTML]{C0C0C0}34 & \cellcolor[HTML]{C0C0C0}30 \\ \hline
	\end{tabular}
\end{table}
\vspace{-0.5cm}

\begin{figure}[!b]
	\vspace{-0.5cm}
	\captionsetup[subfigure]{justification=centering}
	\begin{subfigure}{.238\columnwidth}
		\centering
		\includegraphics[width=\columnwidth]{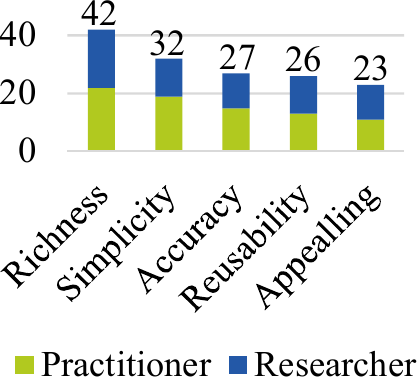}
		\caption{Strengths}
		\label{fig2:sfig1}
	\end{subfigure}
	\rulesep
	\begin{subfigure}{.237\columnwidth}
		\centering
		\includegraphics[width=\columnwidth]{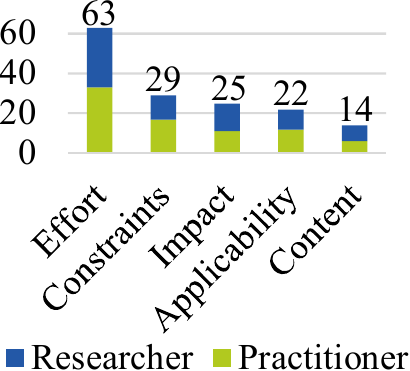}
		\caption{Weaknesses}
		\label{fig2:sfig2}
	\end{subfigure}
	\rulesep
	\begin{subfigure}{.237\columnwidth}
		\centering
		\includegraphics[width=\columnwidth]{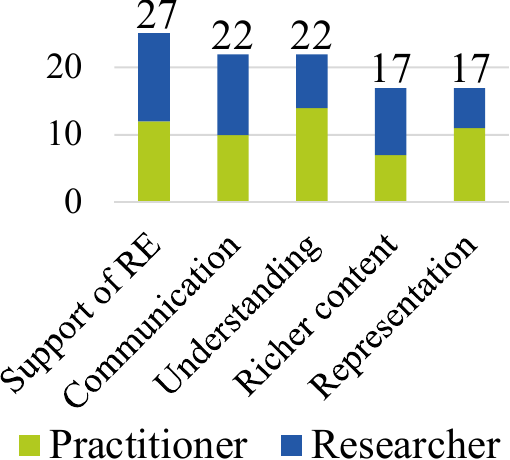}
		\caption{Opportunities}
		\label{fig2:sfig3}
	\end{subfigure}
	\rulesep
	\begin{subfigure}{.237\columnwidth}
		\centering
		\includegraphics[width=\columnwidth]{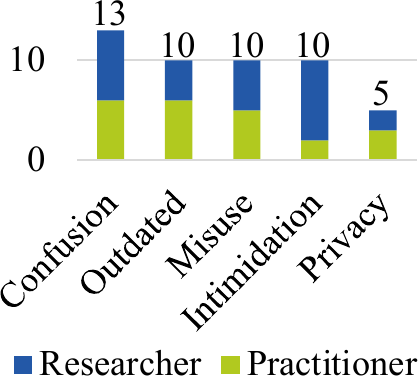}
		\caption{Threats}
		\label{fig2:sfig4}
	\end{subfigure}
	\centering
	\caption{Coding Frequencies (CF) of the open-ended questions}
	\label{fig:fig2}
\end{figure}

\figurename{ \ref{fig:fig2}} shows the Coding Frequencies (CF) for the open-ended questions about the strengths, weaknesses, opportunities, and threats of videos in RE.

\textbf{Strengths:}
Videos are most appreciated for their \textit{richness} (CF: $42$) of detailed and comprehensive information such as gestures, facial expressions, emotions, and rationales. This information can be used and understood fast and easily due to the \textit{simplicity} (CF: $32$) of videos. The respondents emphasized the \textit{accuracy} (CF: $27$) of videos since they capture exact statements and visualize concrete examples, problems, and solutions. Videos have an increased \textit{reusability} (CF: $26$) for later analyses or sharing due to their long-term accessibility and persistence. The visualization of videos is less ambiguous than textual descriptions wherefore videos are an \textit{appealing} (CF: $23$) medium.

\textbf{Weaknesses:}
The most mentioned weakness of videos is the high \textit{effort} \mbox{(CF: $63$)} in terms of costs and time for planning, producing, watching and processing a video. The technical \textit{constraints} (CF: $29$) of videos such as file format, size, or required equipment are a further problem. Videos may have a negative \textit{impact} (CF: $25$) on people with different effects, e.g. too high expectations, intimidation, or low acceptance. The respondents also stated the \textit{applicability} \mbox{(CF: $22$)} of videos as difficult. Besides legal and privacy issues, videos are not suitable for every kind of content and context. Additionally, the information \textit{content} (CF: $14$) of videos is difficult since a video needs to include the right amount of detailed and relevant information.

\textbf{Opportunities:}
The most mentioned opportunity of video is the \textit{support of RE} (CF: $27$) in terms of improving activities (elicitation, interpretation, validation, and documentation) and techniques (interview, workshop, focus group, and observation). Especially, the respondents think that videos can improve \textit{communication} (CF: $22$) and \textit{understanding} (CF: $22$) of all involved parties in RE. According to our respondents, videos can provide a \textit{richer content} (CF: $17$) than textual descriptions due to their increased information content with more detailed and comprehensive information. Videos also allow an improved \textit{representation} (CF: $17$) of workflows, interactions, environments, and scenarios due to a better description by visualization.

\textbf{Threats:}
The most mentioned threat of videos is their \textit{confusion} (CF: $13$) since they contain a lot of unstructured data. Thus, it is challenging to identify the right, important, and meaningful content. The management of videos is also cumbersome since frequent changes are difficult to handle and can easily lead to \textit{outdated} (CF: $10$) information. The \textit{misuse} (CF: $10$) of videos is a further threat since they should not be used as a single medium to convey information. Videos may cause \textit{intimidation} (CF: $10$) of people. The respondents stated that the use of video can lead to a changed behavior and untrue statements by persons who feel uncomfortable or do not want to appear in a video. Some respondents mentioned \textit{privacy} (CF: $5$) concerns with respect to the misuse of recorded information or the violation of privacy.

\subsection{Threats to Validity}

\textbf{Construct Validity:}
The single use of a questionnaire causes a mono-method bias. All collected data is based on a single source and thus only allows restricted explanations of our findings. The respondents' rationales and thoughts behind their answers remain unknown. The findings might also be affected subjectively since the author performed the coding and analysis on his own. This threat was mitigated by using \textit{in vivo coding} to adhere closely to the respondents' actual language found in the qualitative data. We published the questionnaire and all collected data online to increase the transparency of our results \cite{Karras.2018a}.

\textbf{External Validity:}
According to the respondents, all of them were software professionals from industry and academia. Thus, we expect that they belong to the target population. The survey, however, was accessible to anyone to achieve heterogeneity in the respondents' attitudes. We also cannot foreclose that respondents made false statements. However, there was no financial reward and thus little incentive to participate in the survey without giving honest answers.

\textbf{Internal Validity:}
Two important threats to internal validity are maturation and instrumentation. The time taken to complete the survey is crucial. In case of too many questions, respondents may be affected negatively and abort. We refined carefully the questionnaire design to improve the instrumentation (see section \ref{sec:survey----video-as-a-medium-in-re}). In case of an abort, all entered data was deleted to increase the respondents' trust in our research.

\textbf{Conclusion Validity:}
The validity of any scientific evaluation highly depends on the reliability of measures. A good question wording, instrumentation, and instrumentation layout are crucial for the results of a survey. We followed survey guidelines and used LimeSurvey, which is a professional survey software, to ensure these aspects. We consciously decided on the respondents' heterogeneity to increase the external validity. However, the variation in knowledge and background might affect the findings and thus restricts the conclusion validity.

\section{Discussion}
\label{sec:discussion}
We investigate the assumption that software professionals' perceive video as a medium that can contribute to RE. Despite $35$ years of research on integrating videos in RE, this medium is still not an established documentation option. We focus on the attitudes of software professionals towards video to achieve insights whether they fundamentally reject videos in RE or not. Our findings substantiate the assumption, but also indicate crucial concerns in terms of weaknesses and threats that impede the application of videos in RE.

Software professionals generally have a positive attitude towards videos. $59$ out of $64$ respondents stated that videos have the potential to improve RE by supporting multiple RE activities and techniques and by providing a richer content as well as a better representation than textual descriptions. The mentioned strengths of videos (\textit{richness}, \textit{simplicity}, \textit{accuracy}, \textit{reusability}, and \textit{appealing}ness) underline the benefits of video as a powerful, simple, and appealing documentation option. Especially, the top-$3$ opportunities (\textit{support of RE}, improved \textit{communication}, and improved \textit{understanding}) emphasize the suitability of video as a medium in RE for requirements communication.

However, the respondents stated weaknesses and threats that are crucial concerns which impede the application of videos in RE. Especially, the perceived high \textit{effort} to plan, produce, watch, and process videos is the most frequently identified code overall. Besides this primary weakness, further mentioned weaknesses and threats of videos are i.a. technical \textit{constraints}, negative \textit{impact}, \textit{misuse}, and an improper information \textit{content}. All of them indicate a lack of knowledge of software professionals on how to produce and use good videos that are suitable for RE. As an answer to our research question, we can summarize:

\begin{mdframed}
	\begin{itemize}[leftmargin=-2.5mm]
		\item[] \textbf{Answer:}
		Software professionals' attitudes towards video as a medium in RE are mostly positive. They do not fundamentally reject videos in RE. However, besides clear strengths and opportunities, there are crucial weaknesses and threats that impede the application of videos in RE.
	\end{itemize}
\end{mdframed}

These findings coincide with the conclusions of different researchers \cite{Owens.2011, Carter.2009} and the argumentation in our position paper \cite{Karras.2018}. ``The important thing is to know how to visually communicate'' \cite[p. 80]{Owens.2011}. Previous approaches focused on the use of videos in RE but omitted the details about how to produce them \cite{Karras.2018}. So far, little research encountered the challenge of enabling software professionals with the required knowledge to produce and use good videos for visual communication \cite{Karras.2018}. This emphasizes the need for research that focuses on the production of effective videos to establish them as a communication tool in RE practice \cite{Carter.2009}.

We want to encounter this challenge of enabling software professionals to produce and use good videos on their own at moderate costs, yet sufficient quality. For this, software professionals need to understand what constitutes the quality of a good video.
As proposed in our position paper \cite{Karras.2018}, our future work focuses on developing a quality model for videos since such a model allows ($1$) to evaluate the quality of existing videos and ($2$) to guide the video production and use process. Software professionals can use this quality model as an orientation for planning, shooting, post-processing, and viewing videos in RE.

\section{Conclusion}
\label{sec:conclusion}
Despite its low communication richness and effectiveness, a written specification is the most common medium for requirements communication. In contrast, videos allow a richer knowledge transfer. Although several researchers suggested applying videos in RE by proposing corresponding approaches, this medium is still not an established documentation option. We conducted a survey to explore software professionals' attitudes towards video as a medium in RE in order to achieve insights if they fundamentally reject videos.

Based on our findings, software professionals do not fundamentally reject videos. However, this medium still has a neglected potential. The identified strengths and opportunities underline the benefits of video as a documentation option for communication. Nevertheless, videos are also associated with multiple weaknesses and threats that impede their application in RE. We consider our findings as further indicators that substantiate a lack of knowledge of software professionals on how to produce and use good videos for visual communication. Thus, we follow our proposed future work of developing a quality model for videos to encourage and enable software professionals to produce and use effective videos for RE on their own at moderate costs, yet sufficient quality.

\section*{Acknowledgment}
This work was supported by the Deutsche Forschungsgemeinschaft (DFG) under ViViReq, Grant No.:~289386339, (2017 -- 2019).

\bibliographystyle{splncs04}
\bibliography{refs}

\end{document}